# GS-TG: 3D <u>G</u>aussian <u>S</u>platting Accelerator with <u>T</u>ile <u>G</u>rouping for Reducing Redundant Sorting while Preserving Rasterization Efficiency

Joongho Jo
Korea University
Seoul, Republic of Korea
jojoss1004@korea.ac.kr

Jongsun Park
Korea University
Seoul, Republic of Korea
jongsun@korea.ac.kr

*Abstract*—3D Gaussian Splatting (3D-GS) has emerged as a promising alternative to neural radiance fields (NeRF) as it offers high speed as well as high image quality in novel view synthesis. Despite these advancements, 3D-GS still struggles to meet the frames per second (FPS) demands of real-time applications. In this paper, we introduce GS-TG, a tile-grouping-based accelerator that enhances 3D-GS rendering speed by reducing redundant sorting operations and preserving rasterization efficiency. GS-TG addresses a critical trade-off issue in 3D-GS rendering: increasing the tile size effectively reduces redundant sorting operations, but it concurrently increases unnecessary rasterization computations. So, during sorting of the proposed approach, GS-TG groups small tiles (for making large tiles) to share sorting operations across tiles within each group, significantly reducing redundant computations. During rasterization, a bitmask assigned to each Gaussian identifies relevant small tiles, to enable efficient sharing of sorting results. Consequently, GS-TG enables sorting to be performed as if a large tile size is used by grouping tiles during the sorting stage, while allowing rasterization to proceed with the original small tiles by using bitmasks in the rasterization stage. GS-TG is a lossless method requiring no retraining or fine-tuning and it can be seamlessly integrated with previous 3D-GS optimization techniques. Experimental results show that GS-TG achieves an average speed-up of 1.54 times over state-of-the-art 3D-GS accelerators.

## I. INTRODUCTION

3D Gaussian Splatting (3D-GS) [1] is emerging as a promising alternative to neural radiance fields (NeRF) [2], [3], [11], [12], [13] in the field of novel view synthesis. While NeRF implicitly models a scene using neural networks and synthesizes images through neural network inference, 3D-GS explicitly models a scene using millions of 3D Gaussians with learnable properties, and synthesizes images through tile-based rendering. This approach avoids the time-consuming sampling process of NeRF and enables parallel processing for faster rendering. Although 3D-GS has achieved significant speed-ups over NeRF, it still faces challenges in achieving the frames per second (FPS) required by 3D vision applications, such as augmented reality (AR) and virtual reality (VR) devices. For instance, even with Nvidia's server-grade A6000 GPU, the original 3D-GS [1] achieves only 15-25 FPS depending on the scene when synthesizing images at a resolution of 3840 x 2160. This falls short of the 90-120 FPS required for real-time interaction in binocular displays with a resolution of 2x(2014x2208), as required by recent AR/VR devices like Meta Quest 3.

Many previous works aiming to enhance the rendering speed of 3D-GS have applied well-known compression techniques such as quantization [6], densification [5], vector grouping [4], and pruning [6]. While these techniques intuitively reduce computational load, retraining or fine-tuning processes are required to compensate for the degradation in rendering quality. In contrast, algorithmic optimization efforts that improve the efficiency of the rendering pipeline itself without sacrificing rendering quality have been relatively scarce. Recently, GSCore [7] propose techniques that skip unnecessary computations in the tile selection and rasterization stages by considering the shape of Gaussians, enhancing the efficiency of the rendering pipeline. FlashGS [8] presents a method that more precisely recognizes the size and shape of Gaussians for tile selection than GSCore, and it proposes a workflow that operates efficiently running on GPUs.

In the tile-based rendering of 3D-GS, sorting and rasterization operations are performed independently for each tile. So, it inevitably leads to redundant sorting computations and unnecessary rasterization for Gaussians that affect multiple tiles. Our analysis reveals an interesting trade-off: increasing the tile size reduces redundant preprocessing and sorting operations but increases unnecessary rasterization computations. Previous optimization algorithms for 3D-GS rendering, however, have selected the tile sizes based solely on empirical observations of the fastest rendering speeds without thoroughly analyzing this trade-off. Furthermore, in-depth analyses and solutions for addressing this trade-off within the 3D-GS rendering pipeline have never been explored.

In this paper, we introduce GS-TG, a tile-grouping-based accelerator designed to improve the 3D-GS rendering speed by reducing redundant sorting operations while maintaining rasterization efficiency. By grouping tiles, GS-TG shares sorting operations across tiles within each group, significantly reducing redundant computations. During rasterization, a bitmask assigned to each Gaussian identifies its relevant tiles, enabling efficient reuse of sorting results across tiles. One of the main advantages of GS-TG is that it enhances 3D-GS rendering speed by enabling large tile sizes for sorting and smaller tile sizes for rasterization. Additionally, GS-TG is a completely lossless technique, requiring no retraining or fine-tuning, and it can be seamlessly integrated with previous 3D-GS rendering optimization methods.

## II. BACKGROUND

In this section, we introduce the tile-based rendering technique used in 3D Gaussian Splatting (3D-GS) and explain how this method can be applied within the 3D-GS rendering process.

### A. Tile-Based Rendering

Tile-based rendering [17] is an approach that divides the output image into multiple small tiles and performs rendering for each tile individually. This method first identifies the objects that affect the pixel colors of each tile and then proceeds with rasterization on a per-tile basis. Such an approach offers several advantages, including reducing memory bandwidth, supporting parallel processing, and decreasing rasterization operations per pixel.

### B. 3D Gaussian Splatting

3D Gaussian Splatting (3D-GS) [1] is an emerging technique in the field of novel view synthesis that explicitly models a scene using millions of learnable Gaussians and synthesizes images through tile-based rendering. Fig. 1 illustrates the overall rendering pipeline of 3D-GS. This pipeline consists of three primary stages executed sequentially: preprocessing, tile-wise sorting, and tile-wise rasterization.

In the **preprocessing stage**, as shown on the left side of Fig. 1, inputs such as the center position of 3D Gaussians (*3D_XYZ*), 3D

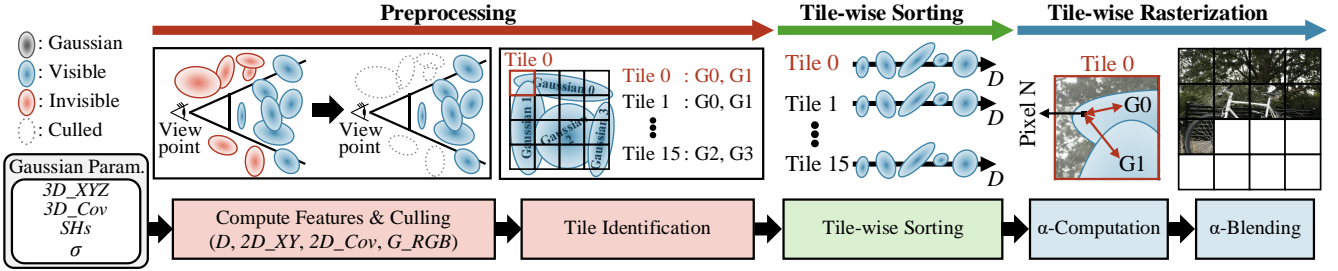

Fig. 1: The overall rendering pipeline of 3D Gaussian Splatting (3D-GS).

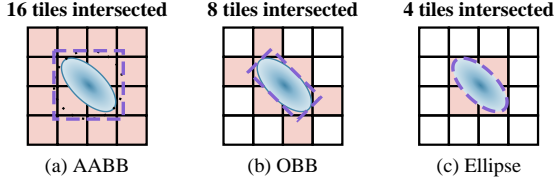

Fig. 2: Comparison of methods for calculating Gaussian influence based on size and shape: (a) AABB, (b) OBB, and (c) Ellipse boundary.

covariance matrix (*3D_Cov*), opacity ($\sigma$), and spherical harmonics coefficients (*SHs*) representing color across different viewpoints are utilized. During this stage, features required for culling and subsequent operations, including depth (*D*), 2D coordinates (*2D_XY*), 2D covariance (*2D_Cov*), and Gaussian color (*G_RGB*), are computed. Concurrently, invisible Gaussians for the current viewpoint are identified and excluded from rendering computations through a culling process. As illustrated in Fig. 1, the initial input includes both visible (blue) and invisible (red) Gaussians, but through culling operations, only Gaussians visible to the viewpoint are used in subsequent computations. Tile identification step, which identifies the Gaussians affecting each tile, uses *2D_XY* and *2D_Cov* to determine the tiles influenced by each Gaussian. While theoretically, the influence range of a Gaussian is infinite, the 3-sigma rule is applied to define the Gaussian's size in [1]. In the **tile-wise sorting stage**, for each tile, the Gaussians identified during preprocessing are sorted in the order of proximity to the viewpoint by utilizing their depth values (*D*). Finally, in the **tile-wise rasterization stage**, the sorted sequence of Gaussians for each tile is utilized to compute pixel colors in the order of proximity to the current viewpoint. As shown in Fig.1, rasterization consists of two main steps. The first step, known as $\alpha$-computation, uses *2D_XY*, *2D_Cov*, and $\sigma$ to calculate how much each Gaussian influences the color of the pixel being processed. Then, $\alpha$-computation for $i$-th Gaussian is calculated using the following equation:

$$\alpha_i = \sigma_i * \exp\left(-\frac{1}{2}(P - 2D\_XY_i)^T 2D\_Cov_i^{-1}(P - 2D\_XY_i)\right), \quad (1)$$

where $P$ means the pixel's coordinates, and $2D\_Cov_i^{-1}$ is the inverse of $i$-th Gaussian's 2D covariance matrix. In [1], when $\alpha_i$ is less than 1/255, $i$-th Gaussian is considered to have no influence on the pixel color and is therefore excluded from the subsequent $\alpha$-blending computation. In $\alpha$-blending step, pixel colors are computed starting from the one closest to the viewpoint, using the following equation:

$$Pixel\ Color = \sum_{i=1}^{N} G\_RGB_i \alpha_i \prod_{k=1}^{i-1}(1 - \alpha_k), \quad (2)$$

where $N$ is the number of visible Gaussian associated with the pixel. During the accumulation operation in (2), an early-exit is triggered if $\prod_{k=1}^{i-1}(1-\alpha_k)$ falls below a predefined threshold (e.g., $10^{-4}$ as suggested in [1]).

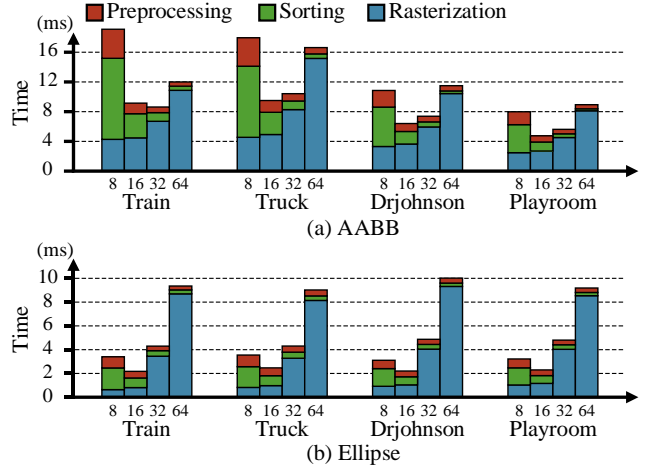

Fig. 3: Runtime breakdown across tile sizes for four scenes at different boundaries: (a) AABB and (b) Ellipse boundary.

*C. Related Works*

Recent advances, such as GSCore [7] and FlashGS [8], have improved 3D-GS rendering by introducing techniques to accurately identify the tiles influenced by Gaussians. These methods consider the size and shape of each Gaussian, reducing the number of Gaussians required for tile-wise sorting and rasterization. Fig. 2 illustrates three methods for calculating the influence of Gaussians. Fig. 2 (a) depicts an example of axis-aligned bounding boxes (AABB), where the Gaussian intersects with 16 tiles when using AABB. This approach, adopted in the original 3D-GS, is computationally efficient but fails to account for the shape of the Gaussian, resulting in significant unnecessary computations. Fig. 2 (b) demonstrates the use of oriented bounding boxes (OBB), where the Gaussian is identified to intersect with 8 tiles. This method, employed in GSCore, reduces unnecessary tile identification compared to AABB but comes at a relatively higher computational cost. Fig. 2 (c) presents the ellipse-based boundary, where the Gaussian intersects with only 4 tiles. This approach, adopted in FlashGS, minimizes unnecessary tile selection, but it is computationally more expensive as it directly handles the elliptical shape rather than a rectangular boundary. However, these works primarily focus on considering the size and shape of Gaussians, without conducting an in-depth analysis of the impact of *tile size* across the stages of the rendering pipeline. In the following section, we present the importance of *tile size* across the stages of the rendering pipeline.

III. MOTIVATION

In this section, we first profile the 3D-GS rendering pipeline under various tile sizes. Subsequently, the impact of tile size on each stage of the 3D-GS rendering pipeline is analyzed.

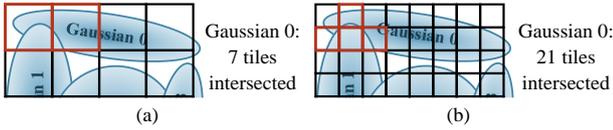

Fig. 4: Tile identification examples for (a) large and (b) small tile sizes.

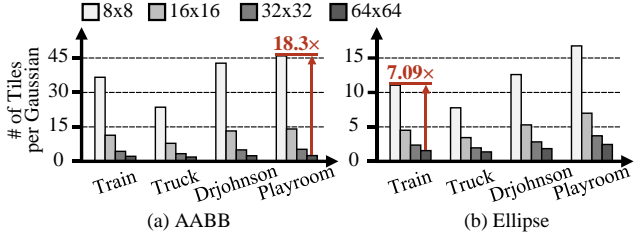

Fig. 5: Average number of intersecting tiles per Gaussian for various tile sizes in (a) AABB and (b) Ellipse boundary across four scenes.

TABLE I. THE PERCENTAGE OF GAUSSIANS SHARED WITH ADJACENT TILES BASED ON TILE SIZE

| % | 8×8 | 16×16 | 32×32 | 64×64 |
|---|---|---|---|---|
| Train | 94.4 | 89.0 | 79.7 | 66.0 |
| Truck | 89.0 | 79.2 | 64.7 | 47.7 |
| Drjohnson | 91.4 | 83.9 | 71.3 | 54.0 |
| Playroom | 91.3 | 83.8 | 71.7 | 54.7 |
| Average | 91.5 | 84.0 | 71.9 | 55.6 |

### A. Runtime Trade-offs in 3D-GS Rendering Across Different Tile Sizes

To analyze the rendering performance of 3D-GS with different tile sizes, profiling is conducted on the NVIDIA A6000. Fig. 3 presents the runtime breakdown across four scenes—Train, Truck, Drjohnson, and Playroom—using various tile sizes (8×8, 16×16, 32×32, and 64×64). The results are shown separately for AABB in Fig. 3(a) and ellipse boundary in Fig. 3(b). For the analysis, we use and modify the author-released codes from [1] and [8]. Detailed information on the scenes and pretrained models can be found in section VI-A. As shown in Fig. 3, the runtime for the preprocessing and sorting stages decreases with larger tile sizes, whereas the rasterization stage finishes more quickly with smaller tile sizes. Generally, a tile size of 16×16 provides the fastest rendering speed, though in some cases, 32×32 can also be faster. However, previous optimization algorithms for 3D-GS rendering have selected tile sizes based solely on empirical observations of rendering speed, without a comprehensive analysis of this underlying trade-off. In section III-B, we conduct an in-depth analysis of this trade-off to identify the opportunities for accelerating 3D-GS rendering.

### B. Analysis of the Impact of Tile Size at Each Stage

**Preprocessing stage.** In the preprocessing stage, tile size affects the computations in the 'tile identification' step, which identifies the tiles influenced by each Gaussian. Fig. 4 illustrates the examples of the tile identification process for (a) a large tile size and (b) a small tile size. As shown in the figure, reducing the tile size increases the number of tiles across the same image area, thereby increasing the number of tiles that each Gaussian needs to process during the tile identification step. Consequently, as the tile size decreases, the runtime of the preprocessing stage increases.

**Tile-wise sorting stage.** In Fig. 4 (a), Gaussian 0 intersects with 7 tiles and shares 2 of these tiles (highlighted with a red border) with Gaussian 1. Since sorting is performed on a per-tile basis, Gaussian 0 undergoes 7 sorting operations, 2 of which are redundant with Gaussian 1. In Fig. 4 (b), Gaussian 0 intersects with 21 tiles, sharing 4 tiles with Gaussian 1, resulting in redundant sorting operations. This outcome is inevitable, as reducing tile size increases the number of tiles within the same image size. Fig. 5 shows the average number of intersecting tiles per Gaussian under the same conditions as in Fig. 3. The results demonstrate that as the tile size decreases, the number of intersecting tiles per Gaussian increases exponentially. For the playroom scene at AABB, with an 8×8 tile size, the average number of tiles intersected by each Gaussian is 18.3 times higher than with a 64×64 tile size. Table I presents the percentage of Gaussians shared with adjacent tiles across different tile sizes. Notably, with an 8×8 tile size, an average of 91.5% of Gaussians are shared with adjacent tiles. In the tile-wise sorting stage, our analysis indicates that in 3D-GS rendering, as the tile size decreases, the number of sorting operations per tile increases exponentially, making a significant portion of these sorting operations redundant.

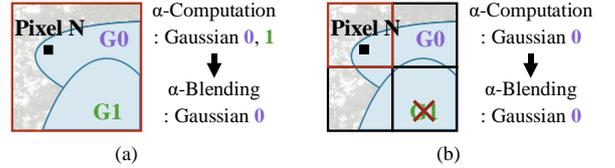

Fig. 6: Tile-wise rasterization examples for (a) large and (b) small tile sizes.

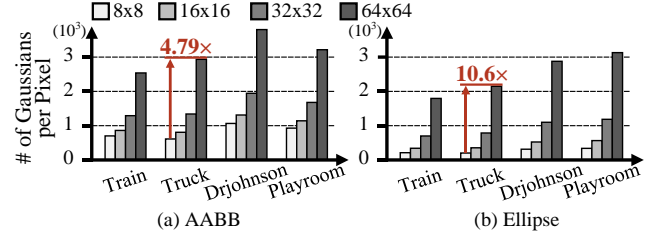

Fig. 7: Average number of Gaussians that need to be processed per pixel in (a) AABB and (b) Ellipse across four scenes.

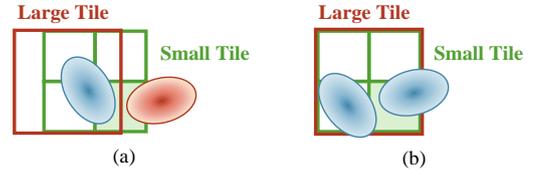

Fig. 8: Scenarios of small tile alignment within a large tile: (a) Misaligned and (b) Perfectly aligned.

**Tile-wise rasterization stage.** Fig. 6 illustrates an example of rasterization with (a) a large tile size and (b) a small tile size. In Fig. 6 (a), the tile containing Pixel N intersects with both Gaussian 0 and Gaussian 1, resulting in rasterization operations for Pixel N with respect to both Gaussians. In this case, Gaussian 0 influences Pixel N, necessitating both $\alpha$-computation and $\alpha$-blending. However, since Gaussian 1 does not affect Pixel N, an $\alpha$-computation is performed to determine its influence, concluding that Gaussian 1 does not impact Pixel N, thereby excluding it from the $\alpha$-blending operation. In Fig. 6 (b), only Gaussian 0 intersects the tile containing Pixel N, so unlike in the large tile size case, no $\alpha$-computation is required for Gaussian 1. As this example shows, reducing the tile size allows for finer identification of Gaussian influence during the preprocessing stage, reducing unnecessary computations in the rasterization stage. Fig. 7 presents the average number of Gaussians that need to be processed per pixel under the same conditions as Fig. 3. The graph reveals that across all scenes and boundaries, as tile size increases, so does the number of Gaussians processed per pixel, indicating an increase in unnecessary $\alpha$-computation operations with larger tile sizes. For the truck scene at ellipse with a 64×64 tile size, the average number of Gaussians that need to be processed per pixel is 10.6 times higher than with an 8×8 tile size. In the tile-wise rasterization operations, our analysis indicates that larger tile sizes lead to an increase in unnecessary rasterization computations.

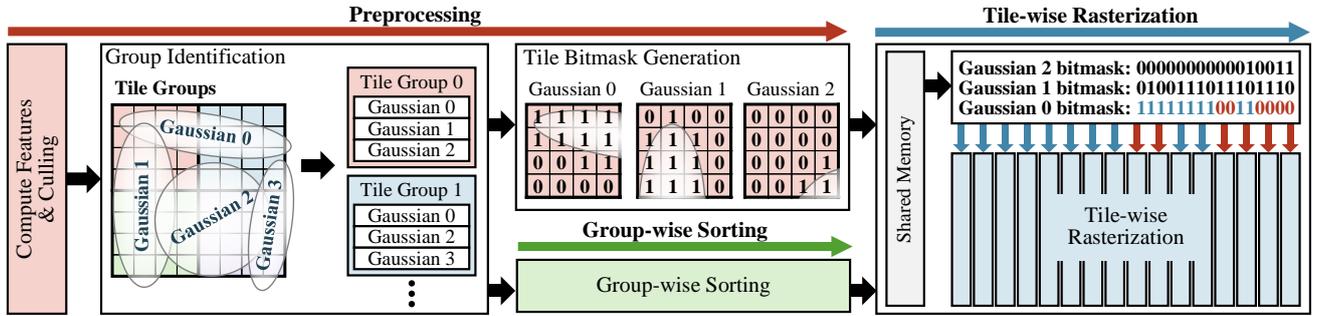

Fig. 9: the overall rendering pipeline of GS-TG.

## IV. PROPOSED TILE-GROUPING-BASED RENDERING PIPELINE

### A. Challenges in Balancing Tile Sizes for Efficient 3D-GS Rendering

Our analysis shows that larger tiles reduce preprocessing and sorting runtime, while smaller tiles improve rasterization efficiency. However, achieving both goals presents two challenges. First, using both small and large tiles requires tile identification operations for both tile sizes during the preprocessing stage, which leads to computational overhead. Second, a method is required to utilize data sorted at the larger tile size for rasterization at the smaller tile size. To address these challenges, we introduce GS-TG, tile-grouping-based rendering pipeline that enhances 3D-GS rendering speed by reducing redundant sorting operations while preserving rasterization efficiency.

### B. Overall Process of GS-TG

To efficiently address both challenges with minimal overhead, smaller tiles should completely fit within larger tiles. Fig. 8 illustrates two scenarios: (a) when smaller tiles do not perfectly align with the larger tile, and (b) when they do. As shown in Figure 8(a), when smaller tiles are not perfectly aligned with the larger tile, some Gaussians (e.g., the red Gaussian in Fig. 8 (a)) may affect only the smaller tiles and not the larger tile. By contrast, as shown in Fig. 8 (b), when smaller tiles are perfectly aligned with the larger tile, all Gaussians that affect the smaller tiles also affect the larger tile. In this way, when smaller tiles fit perfectly within the larger tile, computational independence of the larger tile is ensured. Specifically, once the Gaussians affecting the larger tile are identified, all rendering operations for the smaller tiles within the larger tile can be performed using only those Gaussians. This computational independence provides three key advantages: 1) Even when rendering the smaller tiles using only the Gaussians that affect the larger tile, lossless results are guaranteed. 2) The operation of identifying the Gaussians affecting the smaller tiles and the sorting operation for the larger tile can be performed independently. 3) Gaussians affecting the smaller tiles can be selected using a simple bitmask, eliminating the need for large index overhead. As a result, ensuring that smaller tiles fit perfectly within the larger tile effectively resolves the aforementioned challenges. Therefore, we propose a tile-grouping-based rendering method that treats groups of tiles as if they are larger tiles. This ensures the computational independence of the tile groups (larger tiles).

Fig. 9 illustrates the overall pipeline of tile grouping-based rendering. First, feature calculations and culling operations are performed similarly to conventional 3D-GS rendering.

**Group Identification Step.** Tiles are grouped based on predefined group size, and the Gaussians influencing each group are identified. This step is similar to the tile identification step in conventional 3D-GS rendering but operates at the group level, which corresponds to larger tiles. For instance, in Fig. 9, 16 small tiles are grouped together, and the Gaussians influencing each group are determined. By grouping tiles, we ensure that all the small tiles within a group are influenced only by the Gaussians associated with that group. This relationship allows group-wise sorting and simultaneous bitmask generation to determine which Gaussians influence specific tiles within a group. The runtime overhead of identifying Gaussian influence at the small tile level is effectively hidden by these parallel operations. Furthermore, group-level sorting substantially mitigates redundant computational overhead by minimizing repetitive sorting operations, which are otherwise inherent in small tile-level sorting approaches.

**Bitmask Generation Step.** Once groups are identified, bitmasks are generated to enable tile-wise rasterization. For each Gaussian, a 16-bit bitmask is used to indicate which tiles within a group are influenced. For example, in Fig. 9, bitmasks are created for the Gaussians in Group 0 to identify which of the 16 small tiles are influenced. A tile is marked as "1" in the bitmask if it is affected by the Gaussian. Due to the deterministic relationship wherein tiles within a group are exclusively influenced by the corresponding Gaussians, a 16-bit bitmask efficiently encodes these influences, thereby optimizing both memory representation and computational complexity. This compact encoding substantially reduces the overhead involved in managing Gaussian-tile relationships at smaller tile sizes.

**Tile-wise Rasterization Step.** Using the group-sorted data and generated bitmasks, rasterization is performed at the smaller tile level. This minimizes unnecessary $\alpha$-calculations by limiting rasterization to tiles explicitly marked in the bitmask.

For the computation of Gaussian influence during group identification and tile bitmask generation, any of the following techniques can be applied: AABB, OBB, or ellipse boundary. Fig. 9 demonstrates an example where the ellipse technique is applied.

## V. HARDWARE ARCHITECTURE

### A. Motivation

The proposed GS-TG technique is designed to simultaneously perform bitmask generation and group-wise sorting. So, it enables preprocessing and sorting at the large tile group level while rasterization occurs at the small tile level. However, the single instruction, multiple threads (SIMT) architecture of GPUs presents limitations in parallel execution of bitmask generation and group-wise sorting. As a result, it becomes challenging to hide the execution time of the computations related to the impact of Gaussians on small tiles during the preprocessing stage. To overcome these limitations, a specialized hardware architecture has been designed to efficiently support the proposed rendering pipeline.

### B. Architecture Overview

Fig. 10 illustrates the overall hardware architecture and key block diagram of GS-TG. The architecture consists of two primary modules: preprocessing module (PM) and GS-TG Core, both of which are implemented in four parallel instances to support efficient processing. PM performs Gaussian feature calculation, culling, and group identification operations while maintaining the same structure as the preprocessing stage in conventional 3D-GS rendering. Culled Gaussian features are passed to the GS-TG Core for subsequent stages in the GS-TG rendering pipeline. The GS-TG Core comprises three main hardware modules: bitmask generation module (BGM), group-wise sorting module (GSM), and rasterization module (RM).
**BGM.** This module determines the tiles influenced by each Gaussian using opacity ($\sigma$), projected 2D coordinates (*2D_XY*), and covariance (*2D_Cov*) data. Four tile check units operate in parallel to generate a 16-bit tile bitmask (*Tile_Bitmask*), which is sent to the RM. **GSM.** Using the depth information (*D*) of each Gaussian, this module performs quick sorting. The sorting unit is equipped with 16 comparators, enabling high-speed sorting. The sorted Gaussian indices (*Sorted_G_Idx*) are forwarded to the RM. **RM.** This module is central to efficiently processing tile-based rendering, and it performs the following key operations: 1) Utilizing the 16-bit *Tile_Bitmask* and 16-bit *Tile_Location*, it applies bitwise AND operations to filter Gaussians affecting specific tiles. It also applies bitwise OR operations to generate valid flags, indicating whether a Gaussian impacts the current tile. This process handles eight Gaussians in parallel. 2) Gaussian indices identified for specific tiles are stored in a first-in-first-out (FIFO) buffer to maintain rendering order, and those are subsequently transferred to tile-specific rasterization submodules. Tile-wise rasterization module is composed of 16 parallel rasterization units (RUs). Each RU performs $\alpha$-computation and $\alpha$-blending to calculate final pixel colors, which are then output to memory.

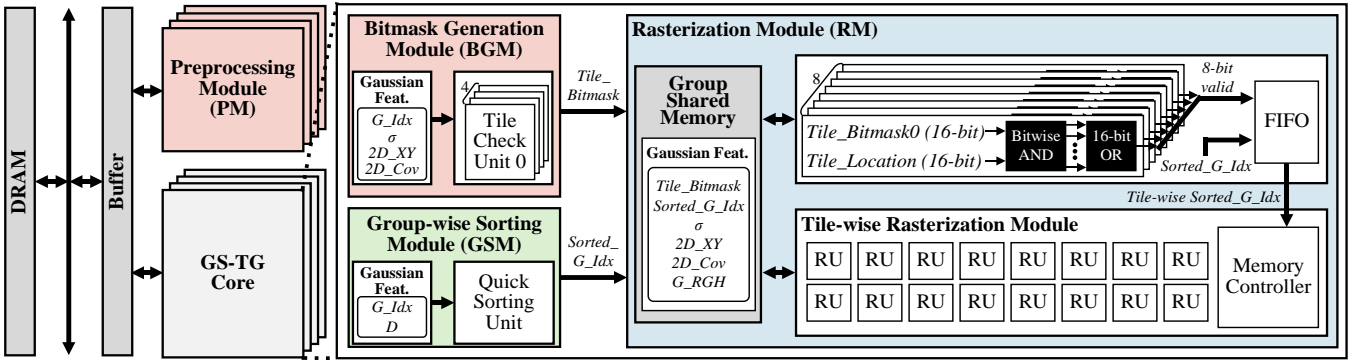

Fig. 10: The overall hardware architecture of GS-TG and the detailed block diagram of each module.

This ensures seamless and efficient parallel processing across all stages, maximizing the performance of the GS-TG rendering pipeline.

## VI. Experimental Results

### A. Methodology

**Algorithm**. As mentioned in Section V-A, due to the single instruction multiple thread (SIMT) architecture of GPUs, it is challenging to execute bitmask generation and group-wise sorting concurrently in parallel. Therefore, GS-TG is designed to operate sequentially in the following order: group identification, tile bitmask generation, group-wise sorting, and tile-wise rasterization—whether running on a GPU. GS-TG is implemented by modifying the author-released codes from [8] using CUDA programming. Experiments are conducted on an NVIDIA A6000 GPU, and the codes are compiled with GCC 11.4.0 and NVCC in CUDA 11.6. **Hardware.** The proposed GS-TG architecture is implemented in RTL and synthesized using Synopsys Design Compiler with 28nm CMOS technology. Power and energy consumption are simulated using Synopsys PrimeTime PX, and speed improvements are evaluated using a cycle-level simulator. The DRAM bandwidth was set to 51.2 GB/s, and DRAM energy consumption is calculated based on [16]. **Common Details.** Table II presents the resolution and types of datasets used for evaluation. From the Tanks&Temples [9], Deep Blending [10], Mill-19 [14], and UrbanScene3D [15] datasets, various scenes representing indoor and outdoor real-world environments are selected: train, truck, Drjohnson, playroom, rubble, and residence. Simulations are performed using pre-trained 3D-GS-30k models. And, to improve the throughput and area efficiency of GS-TG, the models trained in 32-bit floating point are converted to 16-bit floating point. Following the methodology proposed in Mip-NeRF360, a train/test split is applied to all datasets. For the T&T and DB datasets, every 8th image is used for testing. In the Mill-19 dataset, every 64th image is used for testing, while in the UrbanScene3D dataset, every 128th image is used for testing.

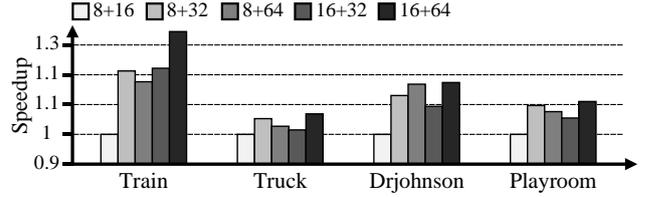

Fig. 11: Speedup of GS-TG with various boundary method combinations across four scenes running on GPU.

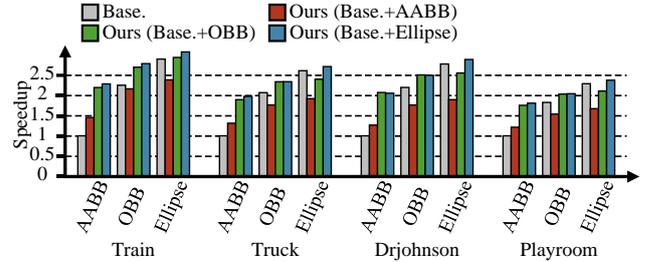

Fig. 12: Speedup of GS-TG with various boundary method combinations across four scenes running on GPU.

TABLE II. Resolution and types of datasets used for evaluation

| Dataset | Scene | Resolution | Type |
|---|---|---|---|
| Tanks&Temples [9] | Train | 1959×1090 | Outdoor |
|  | Truck | 1957×1091 |  |
| Deep Blending [10] | Drjohnson | 1332×876 | Indoor |
|  | Playroom | 1264×832 |  |
| Mill-19 [14] | Rubble | 4608×3456 | Outdoor |
| UrbanScene3D [15] | Residence | 5472×3648 | Outdoor |

### B. Algorithm Evaluation

We first evaluate the performance of various tile sizes and group sizes by considering the parallel operations of bitmask generation and group-wise sorting. Fig. 11 shows the normalized speedup across four scenes (train, truck, Drjohnson, playroom). In Fig. 11, "8+16" indicates a tile size of 8×8 and a group size of 16×16, meaning that each group consists of 4 tiles. According to our analysis, the combination of 16+64 yielded the fastest performance in most cases. Therefore, the subsequent experiments with the proposed GS-TG are conducted with a tile size of 16×16 and a group size of 64×64.

We evaluate the speedup of GS-TG running on a GPU. Fig. 12 illustrates the performance improvements achieved on four scenes: train, truck, Drjohnson, and playroom. The x-axis labels AABB, OBB, and Ellipse represent the boundary methods used in the baseline tile identification step to calculate the influence of

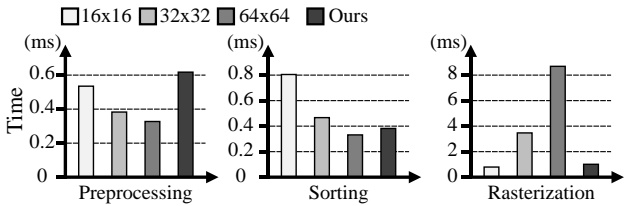

**Fig. 13: Stage-wise runtime breakdown of the rendering pipeline for the Train scene, comparing the baseline approach with GS-TG running on GPU.**

Gaussians. The speed-up results are normalized to the baseline case with AABB applied. In the figure, 'ours' represents the cases where the boundary methods shown on the x-axis are applied in the group identification step instead of the baseline's tile identification step. The red, green, and blue colors represent the cases where AABB, OBB, and Ellipse boundary methods are applied, respectively, in the bitmask generation step of GS-TG. For example, the green bar for AABB indicates the case where AABB is used in the group identification step, while OBB is applied in the bitmask generation step. From Fig. 12, three key findings are observed: 1) When using the Ellipse+Ellipse combination in GS-TG, performance surpassed all the baselines. 2) When the same boundary is applied to both the baseline and GS-TG, GS-TG consistently achieves faster speeds. For instance, GS-TG with OBB+OBB outperformed the baseline using OBB. 3) The proposed tile grouping-based rendering technique demonstrated flexible compatibility with various boundary methods.

Fig. 13 illustrates the stage-wise runtime breakdown of the three rendering pipeline stages (preprocessing, sorting, and rasterization) for the train scene. It compares the baseline, which uses the Ellipse boundary approach with tile sizes of 16×16, 32×32, and 64×64, to the proposed GS-TG employing Ellipse+Ellipse boundary. As shown in the figure, the proposed GS-TG achieves sorting performance comparable to a 64×64 tile size by grouping 16×16 tiles into 16 groups during the sorting stage, while maintaining rasterization performance equivalent to a 16×16 tile size. However, when GS-TG runs on a GPU, the bitmask generation operations cannot overlap with the sorting operations during runtime, causing the preprocessing stage to be slower than the baseline. This issue has been resolved through the dedicated GS-TG accelerator, enabling parallel execution of bitmask generation and group sorting, thereby further enhancing the overall performance.

### C. Hardware Evaluation

Table III presents the synthesis results for each module of GS-TG. The total area of GS-TG is 3.984 $mm^2$, with a power consumption of 1.063W and an operating frequency of 1 GHz. To the best of our knowledge, the only existing hardware related to 3D-GS rendering is GSCore [7]. Therefore, we set the baseline as the conventional 3D-GS rendering using the Ellipse boundary running on the proposed accelerator and compared the performance of GS-TG to GSCore. Fig. 14 illustrates the speedup achieved by GS-TG across six different scenes compared to the baseline and GSCore. GS-TG achieves a geometric mean speedup of 1.33 times over the baseline, with a maximum speedup of 1.58 times in the high-resolution residence scene. Additionally, GS-TG outperforms GSCore by up to 1.54 times in the residence scene. Fig. 15 presents the energy efficiency achieved by GS-TG compared to the baseline and GSCore across six distinct scenes. GS-TG achieves a geometric mean energy efficiency improvement of 2.12 times over the baseline, with a maximum improvement of 2.97 times observed in the high-resolution Residence scene. This superior performance is attributed to the specialized architecture optimized for the tile-grouping-based rendering pipeline.

## VII. CONCLUSION

In this paper, we thoroughly investigated the impact of tile size on the rendering pipeline of 3D Gaussian splatting (3D-GS) and identified key bottlenecks. To address these challenges, we introduced GS-TG, a tile-grouping-based rendering method that significantly enhances the speed of 3D-GS rendering by reducing redundant sorting operations and optimizing rasterization efficiency. GS-TG is a lossless approach, requiring no retraining or fine-tuning, and can be seamlessly integrated with existing 3D-GS optimization techniques. Experimental results show that GS-TG consistently outperforms state-of-the-art 3D-GS hardware, achieving up to 1.54 times speed-up. This work highlights the potential for further hardware-software co-design optimizations in 3D-GS rendering.

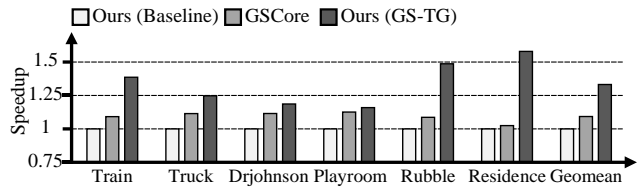

**Fig. 14: Normalized speedup of GS-TG across six different scenes, compared to the baseline and GSCore.**

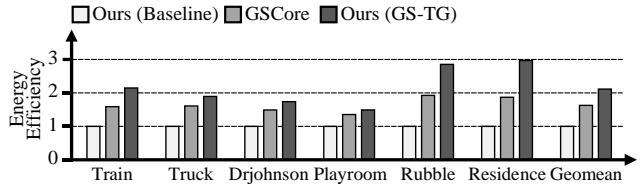

**Fig. 15: Normalized energy efficiency of GS-TG across six different scenes, compared to the baseline and GSCore.**

TABLE III. **HARDWARE CONFIGURATION**

| Module | | Configuration | Area [$mm^2$] | Power [$W$] |
|---|---|---|---|---|
| PM | | 4 | 0.648 | 0.429 |
| GS-TG Core | BGM | 4 | 0.051 | 0.055 |
| | GSM | 4 | 0.012 | 0.001 |
| | RM | 4 | 1.891 | 0.338 |
| Buffer | | 4×2×42KB | 1.382 | 0.240 |
| Total | | | 3.984 | 1.063 |
| Operating Freq. | | | 1GHz | |

ACKNOWLEDGMENT

This work was supported in part by the National Research Foundation of Korea(NRF) grant funded by the Korea government(MSIT) (RS-2024-00345481); in part by the National Research Foundation of Korea(NRF) grant funded by the Korea government(MSIT) (No. RS-2024-00405495, Plug&Play(P&P) Chiplet Integration research center); in part by the Ministry of Trade, Industry and Energy(MOTIE) and Korea Institute for Advancement of Technology(KIAT) through the "International Cooperative R&D program"(Task No. P0028486)